\documentclass[preprint]{elsarticle}

\usepackage{lineno,hyperref}
\usepackage{chemformula} 
\usepackage[T1]{fontenc} 
\usepackage{gensymb}
\usepackage[flushleft]{threeparttable}
\usepackage[normalem]{ulem}
\usepackage{float}
\usepackage{chemscheme}
\usepackage{graphicx}
\modulolinenumbers[5]

\journal{ }

\begin{document}

\begin{frontmatter}

\title{Molecular Description of the Coil-to-Globule Transition of Poly(N-isopropylacrylamide) in Water/Ethanol Mixture at Low Alcohol Concentration}


\author[mymainaddress]{Letizia Tavagnacco}
\author[mymainaddress]{Emanuela Zaccarelli}
\author[mysecondaryaddress]{Ester Chiessi\corref{cor1}}
\cortext[cor1]{Corresponding author}
\ead{ester.chiessi@uniroma2.it}

\address[mymainaddress]{CNR-ISC and Department of Physics, Sapienza University of Rome, Piazzale A. Moro 2, 00185 Rome, Italy}
\address[mysecondaryaddress]{Department of Chemical Sciences and Technologies, University of Rome Tor Vergata, Via della Ricerca Scientifica I, 00133 Rome, Italy}

\begin{abstract}
Poly(N-isopropylacrylamide), PNIPAM, is a widely studied polymer, which serves as a key constituent of nanostructured soft materials with responsive properties. Upon increasing temperature the PNIPAM polymer chain undergoes a reversible coil-to-globule transition at $T\sim$305K, which is reflected by a volume phase transition in cross-linked architectures, such as microgels, valuable for many practical applications. The addition of a cosolvent is a simple method to tune the transition temperature according to the specific purpose. In this study, we use atomistic molecular dynamics simulations to explore the solution behavior of a PNIPAM chain in a mixture of water and ethanol, acting as cosolvent, at low alcohol concentration. Our simulations reproduce the occurrence of the coil-to-globule transition of the polymer chain at 289 K, a temperature lower than that measured in water, in full agreement with experimental findings. By monitoring the temperature evolution of structural and dynamical properties of the PNIPAM-water-ethanol ternary system, we detect a localization of ethanol molecules at the polymer interface, mainly due to interactions between isopropyl and ethyl groups. We observe that the transition occurs without a release of adsorbed ethanol molecules, but with a loss of water molecules from the surrounding of PNIPAM hydrophobic moieties that favours the aggregation of ethanol molecules close to the polymer. Our results support the idea that both the decreased chemical potential of water in the bulk of the mixture and the competition between water and ethanol molecules in the interactions with the polymer play a driving role in the transition.
\end{abstract}

\begin{keyword}
\texttt PNIPAM \sep co-nonsolvency \sep molecular dynamics simulations \sep microgels
\end{keyword}

\end{frontmatter}


\section{Introduction}
Microgels of poly(N-isopropylacrylamide), PNIPAM, are a prototype of the class of stimuli-responsive nanostructured soft-matter systems based on a synthetic covalent network~\cite{fernandez2011microgel}. The temperature modulated water solubility of the PNIPAM polymer, with a lower critical solution temperature (LCST) of about 305 K, originates a volume phase transition in PNIPAM chemically cross-linked micro-hydrogels, that has been extensively studied also for the many chemical or structural variants of these microgels (co-polymers, microgels with uniform or inverted polymer distribution)~\cite{virtanen2016persulfate,mueller2018dynamically,karg2019nanogels,varga2001effect}. Swelled PNIPAM microgels synthesized with the original method~\cite{pelton1986} have a diameter of hundreds of nanometers and display a heterogeneous structure, with a dense core and a looser external region of the particle. The internal nano-structure is suitable to confine molecular entities, such as enzymes or reactants, to promote chemical processes in a selective modality~\cite{fernandez2009gels} and also to prevent ice nucleation~\cite{zanatta2018evidence,tavagnacco2019water}. By exploiting the temperature sensitivity of the polymer chain conformation across the volume phase transition temperature (VPTT), the architecture of the network can be reversibly adapted. Different approaches have been explored to tune the polymer LCST and the related VPTT of PNIPAM based microgels, according to requirement of specific applications. As a result, in the last years several experimental and numerical studies have been carried out to investigate PNIPAM chains and microgels solution behavior~\cite{fernandez2011microgel,rovigatti2019numerical}. The LCST/VPTT control can be achieved by variations of chemical~\cite{kim2019dual} and stereochemical~\cite{biswas2010synthesis,nakano2011thermoreversible,nishi2013} composition of the polymer and by variations of the aqueous suspension medium, concerning ionic strength~\cite{zhang2005,humphreys2016specific,adroher2017conformation} and cosolute~\cite{gao2014,schroer2016} or cosolvent addition, the latter being one of the simplest and less expensive ways to decrease the transition temperature~\cite{wang2014binary,Backes2017,perez2019}. In particular, the presence of ethanol as water cosolvent has a drastic impact on PNIPAM solution behavior, since this polymer, soluble in ethanol at all temperatures and in water at $T\leq$ 305 K, is actually insoluble in ethanol/water mixtures having an ethanol mass fraction approximately from 20\% to 60\%, i.e. a molar fraction from about 0.1 to 0.37, irrespective of temperature~\cite{Hore2013}. Such a behavior, namely the non-solubility of a polymer in the mixture of two liquids that are individually good solvents for it, is named as co-nonsolvency. Moreover, a small addition of ethanol, up to an ethanol molar fraction $x_{ET}$ of 0.1, can lower the polymer LCST of about 15 K~\cite{Trappe2014} and the microgel VPTT of several degrees~\cite{Backes2017}, suggesting a profound rearrangement of the hydration shell of PNIPAM in the presence of ethanol as the minority cosolvent. At higher ethanol concentrations the phase behavior of the polymer mixture is characterized by an upper critical solution temperature (UCST) that depends on both molecular weight and concentration of PNIPAM~\cite{Trappe2014}.

The rationale of co-nonsolvency in PNIPAM-water-cosolvent systems at low cosolvent concentration is an intriguing issue, since the dependence of the phase diagram features on polymer degree of polymerization and concentration defies the commonly accepted rules of polymer solutions~\cite{Kremer2017}. Indeed, the LCST in water-ethanol at low cosolvent concentration is scarcely affected by molecular weight and concentration of the polymer~\cite{Trappe2014}, which is a quite surprising result. The mechanism behind co-nonsolvency has been associated to two views, which mainly focus on: i) the influence of the cosolvent on water affinity for PNIPAM; ii) the specific and selective interaction of the cosolvent with the polymer. For both views, the most widely characterized system is the PNIPAM-water-methanol ternary system. The experimental studies of PNIPAM in water-methanol mixtures of Refs.~\cite{Zhang2001,Sun2010} agree with the first view and relate the thermally anticipated coil-to-globule transition to the formation of different water/methanol complexes, which are poor solvents for PNIPAM. Similarly, the works by Trappe and coworkers~\cite{bischofberger2014,Trappe2014} investigate different ternary systems by scattering and calorimetric measurements, recording that the solubility of PNIPAM in aqueous environment relies on hydration of hydrophobic moieties, which is a water state issue, governed by the energy difference between bulk and shell water. The addition of a neutral cosolvent has the effect of decreasing the water chemical potential in the bulk of the mixture, for the major decrease of molar partial enthalpy, and hence disfavoring polymer hydration. This interpretation is also stated in the simulation study of Bharadwaj et al.~\cite{bharadwaj2018}, where the LCST decrease in the presence of small amounts of cosolvent is investigated by explicit solvent coarse-grained simulations and mean-field theory. Accordingly, a recent experimental and simulation work on poly(N-diethylacrylamide) co-nonsolvency in water/ethanol highlights how water/cosolvent attraction induces phase separation~\cite{zuo2019}.

The second view finds in a specific cosolvent-polymer interaction the major cause of PNIPAM co-nonsolvency. On the basis of a thermodynamical investigation, Schild at al. proposed that local polymer-cosolvent interactions, modulated by water-cosolvent mixture composition, are driving co-nonsolvency at high water contents~\cite{schild1991}. Several numerical and theoretical studies stress that co-nonsolvency can only be explained by the preferential binding of one of the cosolvent components to the polymer~\cite{mukherji2014,mukherji2015,mukherji2013,mukherji2016}. Moreover, the binding between cosolvent molecules localized on distant PNIPAM residues is thought to induce the formation of segmental loops, that trigger the collapse of the coil structure~\cite{mukherji2014}. Tanaka and co-workers~\cite{tanaka2008,tanaka2011}, by extending their cooperative hydration model~\cite{okada2005} to the water-cosolvent mixture, raised the hypothesis that the competitive hydrogen bonding of the cosolvent with the polymer is the driving force behind co-nonsolvency, again stating a selective adsorption mechanism. The molecular simulation study of PNIPAM-water-methanol solution of Rodr{\'\i}guez-Ropero et al. shows that, at low alcohol content, methanol preferentially binds to the PNIPAM globule and drives polymer collapse, because of the increase of the globule's configurational entropy~\cite{rodriguez2015mechanism}. For the same system, in agreement with Tanaka and co-workers~\cite{tanaka2008,tanaka2011}, Dalgicdir et al. find that the presence of methanol in the solvation shell frustrates the ability of water to form hydrogen bonds with the amide proton of PNIPAM, therefore causing the polymer collapse~\cite{dalgicdir2017}.

However, recent investigations demonstrated that the preferential affinity of the cosolvent toward PNIPAM is not a general requirement for co-nonsolvency~\cite{budkov2014,wang2017}. Furthermore, Pica and Graziano~\cite{pica2016} bring into question that the preferential binding of cosolvent molecules to PNIPAM is the cause of co-nonsolvency in water-methanol. Their theoretical study proposes that polymer chains collapse because the competition between water and methanol molecules to interact with the polymer surface favors the solvation energy of the globule state, as compared to the coil conformation, due to geometric frustration. A finding mediating the i) and ii) views is presented in the quasi-elastic neutron scattering study of PNIPAM in water-methanol mixture, where polymer-water, polymer-methanol, and methanol-water hydrogen bonding have been detected~\cite{kyriakos2016}. This evidence proves that the co-nonsolvency phenomenon may be dependent on both solvent-cosolvent and polymer-cosolvent interactions.

With this background, our work aims to obtain a molecular description of the temperature-induced coil-to-globule transition in a water/ethanol solution having a low ethanol content, exploring in particular the evolution of PNIPAM solvation shell as a function of temperature across the LCST. A number of recent theoretical and simulation studies addressed the solution behavior of PNIPAM in water/methanol with reference to the co-nonsolvency phenomenon~\cite{zuo2019,mukherji2013,mukherji2016,tanaka2011,rodriguez2015mechanism,dalgicdir2017,pica2016}, whilst that in water/ethanol is less investigated. In particular, simulations with an atomistic detail of PNIPAM in water/ethanol are missing, with the exception of the work of Backes at al.~\cite{Backes2017}, where the chain conformation behavior in mixtures with ethanol volume fraction of 0.10 and 0.55 is focused. For our solution model we considered the binary mixture with composition $x_{ET} = 0.07$, that corresponds to the minimum of the partial molar volume of ethanol for temperatures lower than 303 K~\cite{armitage1968}. This suggests that at such composition, for a temperature range that includes PNIPAM soluble states and the coil-to-globule transition, occurring at $\sim$290 K~\cite{Trappe2014}, ethanol experiences the maximum steric interaction with water in the binary mixture and the selective adsorption of the alcohol on a solute should be less favored, as compared to other mixture compositions. Interestingly, at $x_{ET} \cong 0.07$ another type of PNIPAM-based systems, such as hydrogels, display the maximum shrinking effect in the whole water-ethanol composition interval at room temperature~\cite{mukae1993swelling}. De facto, the role of a possible selective adsorption of ethanol on PNIPAM is a debated aspect in water-ethanol at low ethanol content, as well as its influence in the reduction of the solubility temperature range of the polymer. The energetic state of the aqueous medium is recognized as a key parameter for PNIPAM phase behavior in the water-rich regime of water-alcohol mixtures~\cite{bharadwaj2018}, which does not imply a preferential interaction of the polymer with the cosolvent. Moreover, the evidence of the LCST independence on the polymer concentration in water-ethanol mixtures is in disagreement with the idea of a major absorption of ethanol~\cite{Trappe2014}. Opposite results are obtained from NMR measurements on PNIPAM gels in binary water/ethanol mixtures which indicate that the polymer preferentially interacts with alcohol molecules over water, in particular for temperatures above the LCST~\cite{wang2009}. In this context, a simulation study allows to relate the conformational features of the polymer chain with the composition, structure and dynamics of the solvation shell. This information, only accessible through an atomistic description of the system, is relevant for understanding the basic principles behind the function of PNIPAM based nano-devices.

In this work we successfully reproduce the coil-to-globule transition, in agreement with experimental behavior, and disclose a modulation of mixture composition in proximity of the polymer as a function of temperature. We detect a significant localization of ethanol molecules at the polymer interface, mainly characterized by interactions with the polymer isopropyl moieties. However, our findings suggest that the driving effect behind the decrease of the coil-to-globule transition temperature, as compared to the pure aqueous solution system, is not the preferential binding of ethanol molecules to the polymer, but rather the decreased chemical potential of water in the bulk of the mixture and the competition between water and ethanol molecules to interact with PNIPAM. On this basis we can state that the two views, that we have described earlier on, do not actually represent alternative scenarios, since factors belonging to both of them characterize the PNIPAM solution behavior in water/ethanol mixture at low alcohol concentration.

\section{Materials and Methods}

\subsection{Model and simulation procedure}

We investigate the solution behavior of a PNIPAM linear chain in dilute regime in a mixture of water and ethanol molecules. We use a binary mixture of solvents with an ethanol molar fraction $x_{ET}=mole_{ethanol}/(mole_{water}+mole_{ethanol})=0.07$. The polymer chain is made of 30 repeating units. This value of degree of polymerization is chosen on the basis of experimental results which suggested that the behavior of an oligomer composed of 28 repeating units is similar to that observed for a polymer with a degree of polymerization of 100~\cite{lutz2006,shan2009}. Consistently, previous simulation studies of PNIPAM 30-mers in water were able to reproduce the coil-to-globule transition of the single chain~\cite{tucker2012,chiessi2016,tavagnacco2018molecular}.

\noindent Tacticity was shown to affect the phase behavior of PNIPAM in water, with a decrease of solubility at higher isotacticity degrees~\cite{suito2002,ito2006,chiessi2016,paradossi2017}. Therefore, to mimic the experimentally measured conditions, we model the polymer chain with an atactic stereochemistry~\cite{chiessi2016,tavagnacco2018molecular}. The configuration of amide groups is also considered by representing their arrangement with a trans geometry.
The polymer is described with the OPLS-AA force field~\cite{jorgensen1996} with the modifications of Siu et al.~\cite{siu2012}. Water and ethanol molecules are modeled with the Tip4p/2005~\cite{tip4p2005} and OPLS-AA~\cite{jorgensen1996} force fields, respectively, a combination of force fields that was shown to properly model the mixture of these solvents~\cite{gereben2015investigation}. The density values of the model solutions are summarized in the Table S1 of the Supplementary Material and are consistent with the experimental values of water-ethanol mixture at $x_{ET} = 0.07$~\cite{density1913}.

\noindent The simulation setup was implemented by centering the polymer chain with an extended conformation in a cubic box of 8.5 nm side and by orienting it along a box diagonal to maximize the distance between periodic images. Then, 1214 ethanol molecules, randomly distributed, and 15416 water molecules were added and an energy minimization with tolerance of 1000 $kJ mol^{-1} nm^{-1}$ was carried out. The resulting system has a polymer mass fraction of 1\% and it was used as initial configuration for the simulations at 7 temperature conditions, i.e. 273 K, 278 K, 288 K, 293 K, 298 K, 308 K and 318 K.
MD simulations were carried out in the NPT ensemble for 350 ns at each temperature. Trajectories were acquired with the leapfrog integration algorithm~\cite{hockney1970potential} with a time step of 2 fs. Cubic periodic boundary conditions and minimum image convention were applied. The length of bonds involving H atoms was constrained by the LINCS procedure~\cite{hess1997lincs}. The velocity rescaling thermostat coupling algorithm, with a time constant of 0.1 ps was used to control temperature~\cite{bussi2007canonical}. The pressure of 1 atm was maintained by the Parrinello-Rahman approach, with a time constant of 2 ps~\cite{parrinello1981polymorphic,nose1983constant}. The cutoff of nonbonded interactions was set to 1 nm and electrostatic interactions were calculated by the smooth particle-mesh Ewald method~\cite{essmann1995smooth}. Typically, the final 40 ns of trajectory were considered for analysis, sampling 1 frame every 5 ps.
Trajectory acquisition and analysis were carried out with the GROMACS software package (version 2018.1)~\cite{markidis2015solving,abraham2015gromacs}. The molecular viewer software package VMD was used for graphic visualization~\cite{humphrey1996vmd}.

\subsection{Trajectory analysis}

The structural properties of the polymer chain were investigated by monitoring the temperature dependence of the radius of gyration ($R_G$) which was calculated through the equation:

\begin{eqnarray}
R_g&=&\sqrt{\frac{\sum_{i}\|r_i\|^2m_i}{\sum_{i}m_i}}
\label{Eq:rg}
\end{eqnarray}

\noindent where $m_i$ is the mass of the $i^{th}$ atom and $r_i$ the position of the $i^{th}$ atom with respect to the center of mass of the polymer chain.

\noindent The solvent accessible surface area (sasa) can be defined as the surface of closest approach of solvent molecules to a solute molecule, where both solute and solvent are described as hard spheres. The sasa can be numerically calculated as the van der Waals envelope of the solute molecule extended by the radius of the solvent sphere about each solute atom centre~\cite{richmond1984solvent}. We used a spherical probe with radius of 0.14 nm and the values of Van der Waals radii of the work of Bondi~\cite{bondi1964van,eisenhaber1995double}. The distributions of sasa values were calculated with a bin of 0.1 nm$^2$.

\noindent The root mean square fluctuation (RMSF) of each backbone carbon atom was estimated as:

\begin{eqnarray}
RMSF_i&=&\sqrt{\langle [\vec{b_i}- \langle \vec{b_i} \rangle ]^2 \rangle}
\label{Eq:RMSF}
\end{eqnarray}

\noindent where $b_i$ and $\langle b_i \rangle $ are the instantaneous and the time averaged position of the backbone atom $i$, respectively. The RMSF was calculated within the time interval of 250 ps and over the whole production run. A roto-translational fit to the structure at the starting time was applied.

\noindent PNIPAM intra-molecular hydrogen bonds (HBs) and HBs with water/ethanol molecules were  monitored as a function of temperature. The occurrence of the hydrogen bonding interaction was evaluated adopting the geometric criteria of an acceptor-donor distance ($A \cdot \cdot \cdot D$) lower than 0.35 nm and an angle $\theta$ ($A \cdot \cdot \cdot D-H$) lower than $30^\circ$, irrespective of the $AD$ pair. The dynamical behavior of hydrogen bonding interactions was characterized by calculating the normalized intermittent time autocorrelation function which is irrespective of intervening interruptions. The corresponding lifetime was defined as the time at which the autocorrelation function is decayed of the 50\% of its amplitude to reduce statistical noise.

\noindent Solvent molecules properties were also investigated by comparing the behavior in the first solvation shell to that in the bulk. The ensemble of water molecules in the first solvation shell was sampled by selecting molecules having the water oxygen atom at a distance from nitrogen, or oxygen, or methyl carbon atoms of PNIPAM lower than the first minimum distance of the corresponding radial distribution function (Figure S1 of the Supplementary Material). Such cutoff values correspond to 0.35 nm for nitrogen and oxygen atoms  and to 0.55 nm for methyl carbon atoms. In the case of ethanol, the cutoff value for methyl carbon atoms was increased to 0.65 nm to account for the hydrophobic groups of the ethanol molecule. Bulk water and ethanol ensembles were evaluated by sampling molecules at a distance larger than 2 nm from the polymer. Clusters of ethanol molecules in the first solvation shell of PNIPAM were calculated using a cutoff distance of 0.35 nm.

\noindent The mobility of first solvation shell solvent molecules was analyzed by selecting water/ethanol molecules at time $t_0$ and then calculating the fraction of molecules that still belongs to the first solvation shell at time $t_0 + t$. The time behavior was evaluated in a time interval of 20 ns averaged over 40 ns with a time between reference points $t_0$ of 100 ps. The corresponding residence time was determined as the time at which the autocorrelation function is decayed of the 50\% of its amplitude to reduce statistical noise.
\noindent The ethanol molar fraction $x_{ET}$ as a function of the distance from the polymer surface was calculated by partitioning the solvent in shells of 0.2 nm up to a distance of 1 nm and of 0.5 nm from 1 nm to 2 nm.

\noindent The diffusion coefficient of water and ethanol molecules was calculated from the long-time slope of the mean square displacement:

\begin{eqnarray}
D&=&\frac{1}{6}\lim_{t\to \infty} {d \over dt} \langle |r(t)-r(0)|^2 \rangle
\label{Eq:Dw}
\end{eqnarray}

\noindent where $r(t)$ and $r(0)$ correspond to the position vector of water/ethanol oxygen atom at time $t$ and $0$, respectively, with an average performed over both time origins and water molecules. To evaluate the limiting slope, a time window of 1 ns was considered.

\section{Results and Discussion}

We provide a description of the simulation results by first focusing the structural features of the polymer and surrounding solvents, then the dynamical properties of PNIPAM, water and ethanol molecules and finally the mechanism of coil-to-globule transition in the binary mixture of solvents.

\subsection{Temperature dependence of polymer structure and solvation}

\begin{figure}[H]
\centering
\includegraphics[width=12cm]{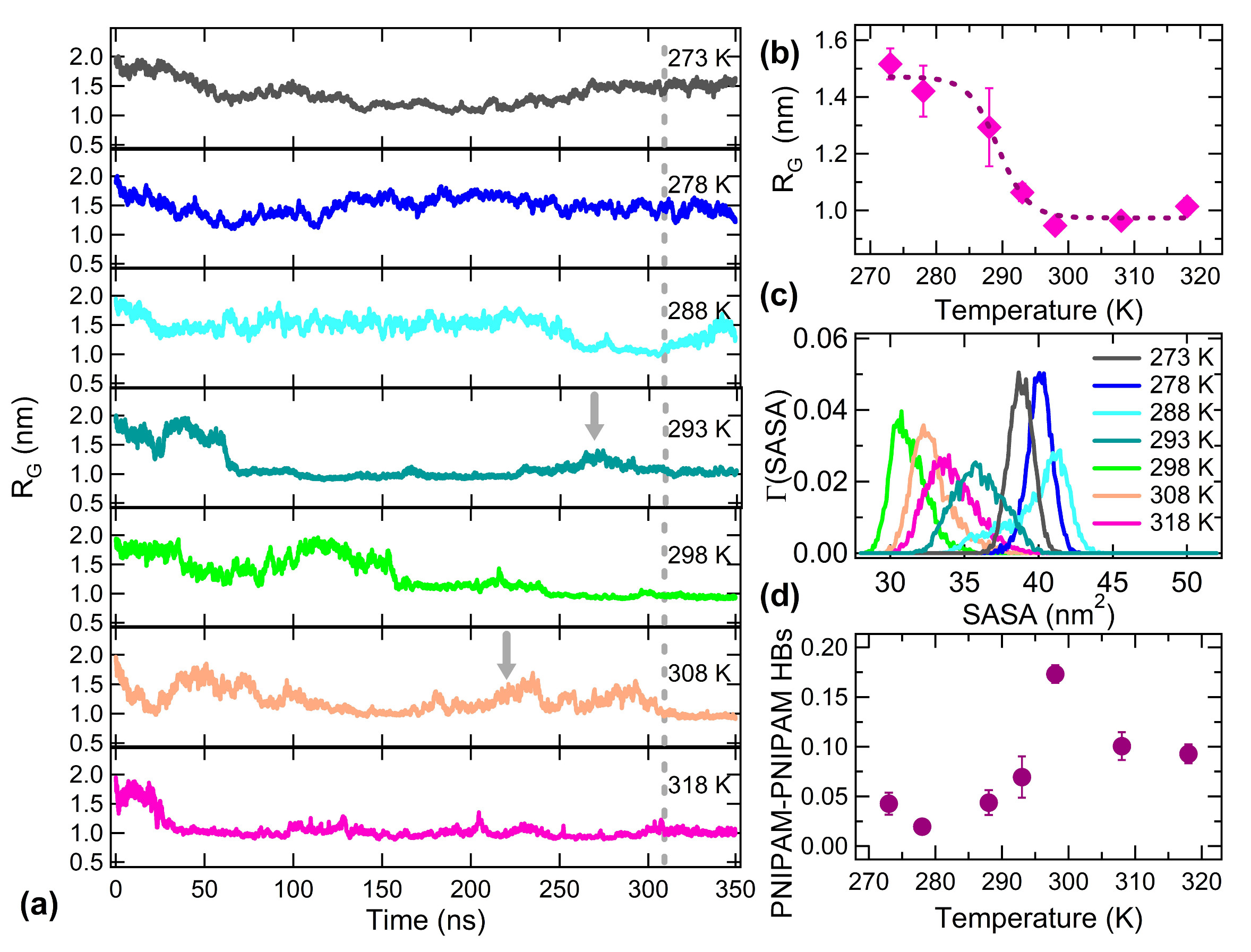}
\caption{(\textbf{a}) Time evolution of PNIPAM radius of gyration as a function of temperature. The vertical dashed line marks the time interval used for the calculation of averaged properties. Gray arrows show recoiling events subsequently the collapse of the chain. (\textbf{b}) Temperature dependence of PNIPAM radius of gyration averaged over the last 40 ns of simulation. The dashed line is the sigmoidal fit. (\textbf{c}) Distributions of values of solvent accessible surface area as a function of temperature (273 K, 278 K, 288 K, 293 K, 298 K, 308 K, and 318 K are shown in dark gray, blue, cyan, dark green, green, orange and magenta, respectively) calculated over the last 40 ns of trajectory. (\textbf{d}) Average number of PNIPAM intra-molecular hydrogen bonds as a function of temperature normalized to the number of repeating units in the polymer chain. When not distinguishable, error bars are within symbol size.}
  \label{fgr:pnipam}
\end{figure}

The occurrence of the coil-to-globule transition of the PNIPAM linear chain in water/ethanol mixture was monitored by calculating the radius of gyration of the polymer as a function of the simulation time, as displayed in Figure~\ref{fgr:pnipam}(a) for the different temperatures investigated. The time evolution of the radius of gyration reveals that at the lowest temperatures of 273 K and 278 K the polymer chain mostly assumes coil conformations, whereas at 288 K significant conformational rearrangements take place and at temperatures higher than 293 K the globule conformations are dominant. At the highest temperatures which are associated to the insoluble states, processes of recoiling of the polymer are detected subsequently the collapse of the chain, as it has been reported in previous studies in aqueous solution~\cite{kang2016collapse,tavagnacco2018molecular}. The temperature dependence of the averaged value of the radius of gyration is displayed in Figure~\ref{fgr:pnipam}(b). The sigmoidal profile of the temperature behavior of the radius of gyration clearly marks the occurrence of the coil-to-globule transition at a temperature of 289($\pm1$) K (Figure~\ref{fgr:pnipam}(b)). The value of transition temperature detected in our simulations coincides with the experimental LCST of PNIPAM in the same binary mixture~\cite{Trappe2014} and is lower than the corresponding transition temperature observed in PNIPAM aqueous solutions~\cite{tucker2012,chiessi2016,tavagnacco2018molecular}. However, the average radius of gyration of the polymer chain in the coil and globule states in water/ethanol solutions is comparable to that observed in aqueous solution~\cite{tavagnacco2018molecular}, with values of about 1.5 nm and 1 nm, respectively.

Another observable which is suitable to probe the conformational transition is the solvent accessible surface area. The general trend of the distributions of sasa values reported in Figure~\ref{fgr:pnipam}(c) shows the presence of two distinct effects of temperature: i) an increase of temperature generally favours the population of conformational states with more extended sasa values and ii) a rise of temperature from 288 K to 298 K determines a net contraction of sasa values which can be related to the occurrence of the coil-to-globule transition. In addition, the comparison of the distribution of sasa values as a function of temperature in water/ethanol mixture with those observed in aqueous solution~\cite{tavagnacco2018molecular} reveals that overall the presence of ethanol molecules determines a slight shift of distribution curves towards lower sasa values.

\begin{scheme}[H]
\centering
\includegraphics[width=11 cm]{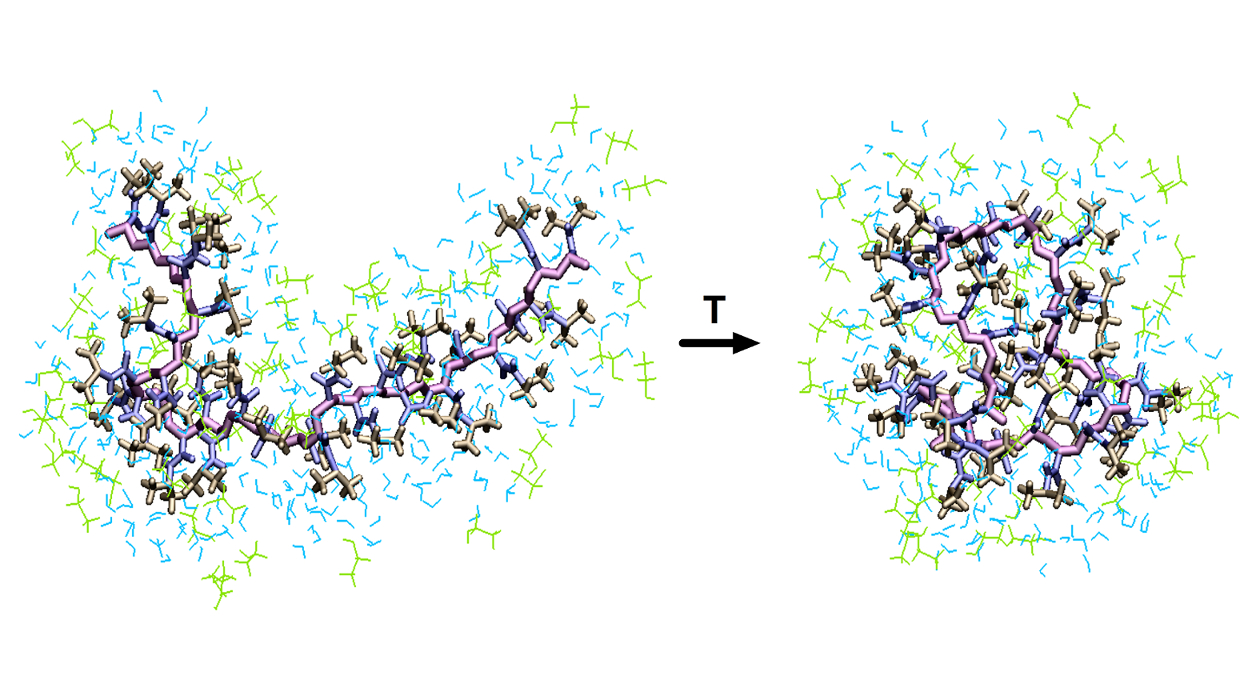}
\caption{Snapshots from the MD simulations at 350 ns showing an example of coil and globule configuration at 278 K (left side) and 298 K (right side), respectively. PNIPAM backbone carbon atoms, side chain hydrophilic groups and side chain hydrophobic groups are represented in pink, blue, and yellow. First solvation shell water and ethanol molecules are displayed in light blue and green, respectively.}
  \label{fgr:scheme}
\end{scheme}

The chain collapse can allow for the formation of PNIPAM intra-molecular hydrogen bonding interactions, although it has been shown that this enthalpic contribution, which favours globule conformational states, is not the main responsible of the coil-to-globule transition~\cite{pica2019does}. Figure~\ref{fgr:pnipam}(d) displays the temperature dependence of the average number of hydrogen bonds between PNIPAM residues normalized to the number of polymer repeating units. In the coil conformational states we observe on average 0.05 HBs per repeating unit, whereas in the globule conformational states the value of HBs doubles. A higher number of PNIPAM intra-chain HBs is found at 298 K which can be related to the formation of more stable globule conformational states, as shown by the time evolution of the radius of gyration (Figure~\ref{fgr:pnipam}(a)). Similar values of intramolecular HBs were determined from simulations in aqueous solutions~\cite{tavagnacco2018molecular}, therefore the presence of ethanol molecules does not significantly affect the formation of intramolecular HBs.

\begin{figure}[H]
\centering
\includegraphics[width=12 cm]{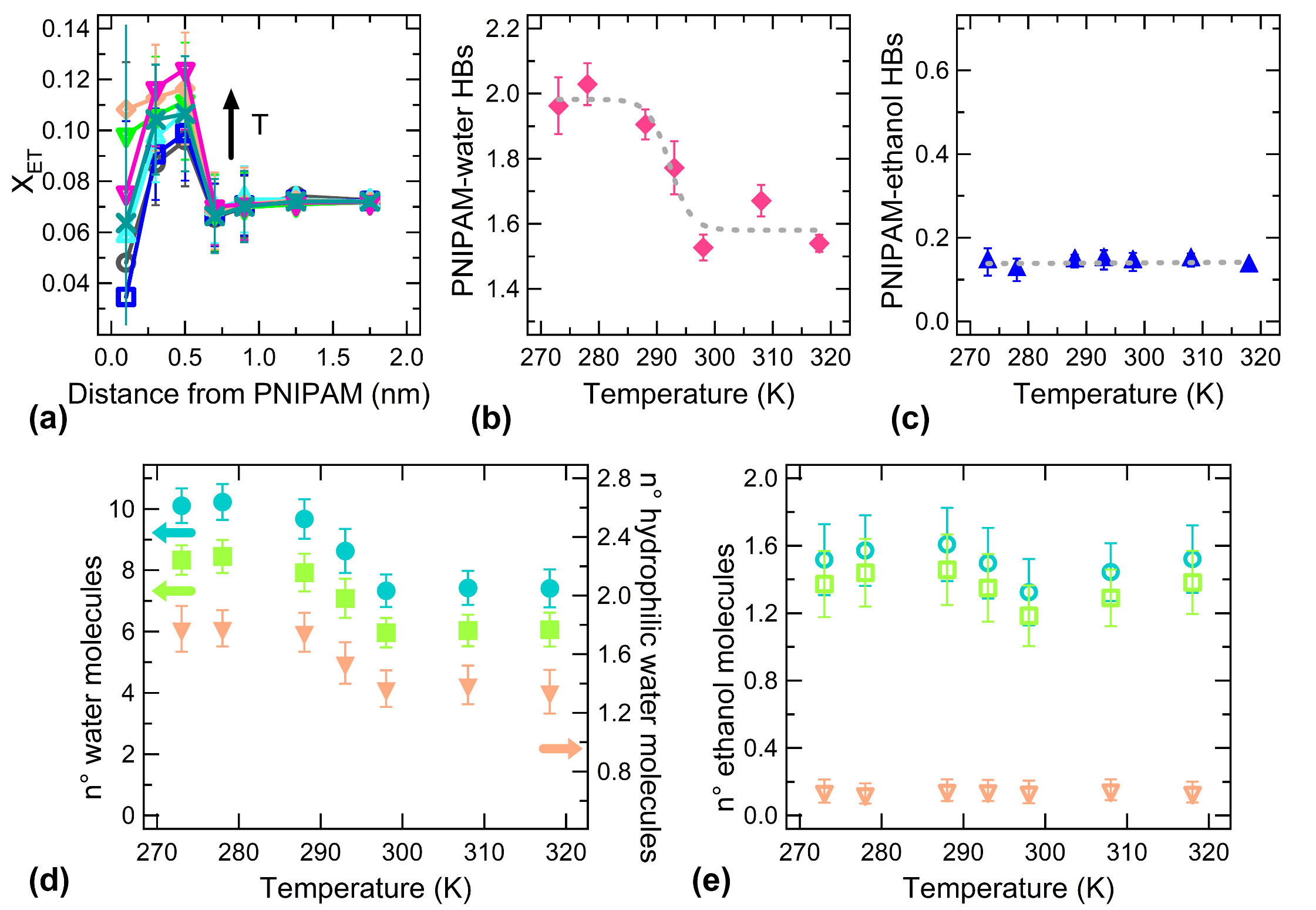}
\caption{(\textbf{a}) Ethanol molar fraction as a function of the distance from PNIPAM surface averaged over the last 40 ns of trajectory. Data calculated at 273 K, 278 K, 288 K, 293 K, 298 K, 308 K, and 318 K are shown with dark gray circles, blue squares, cyan triangles pointing up, dark green crosses, green triangles pointing down, orange diamonds and magenta triangles pointing right, respectively. (\textbf{b}) and (\textbf{c}) Temperature dependence of PNIPAM-water (magenta diamonds) and PNIPAM-ethanol (blue triangles) hydrogen bonds normalized to the number of PNIPAM repeating units. Dotted lines are a guide to the eye. (\textbf{d}) and (\textbf{e}) Average number of first shell water (full symbols) and ethanol (empty symbols) molecules normalized to the number of PNIPAM repeating units and reported as a function of temperature. The total number of solvation molecules is shown with cyan circles, while the number of water molecules interacting with hydrophilic and hydrophobic groups are displayed with orange triangles and green squares, respectively. The arrows indicate the corresponding labelled axis.}
  \label{fgr:solvent}
\end{figure}

Having verified that our simulations successfully reproduce the coil-to-globule transition, we now investigate the organization of solvent molecules surrounding the polymer chain. The inspection of the Scheme~\ref{fgr:scheme}, which illustrates a configuration of the coil and globule states together with the first shell solvation molecules, suggests that ethanol molecules exhibit a good affinity for PNIPAM for all temperatures. To assess such affinity we have probed the ethanol molar fraction $x_{ET}$ as a function of the distance from the polymer chain and of temperature. The results displayed in Figure~\ref{fgr:solvent}(a) show that the concentration of ethanol molecules is considerably higher in the first solvation shell of PNIPAM, as compared to the bulk binary mixture, at all temperatures, and that the extent of ethanol adsorption gradually increases with temperature. This behavior reveals a preferential solvation of the chains by ethanol molecules not only above the transition temperature, as reported by Backes at all.~\cite{Backes2017}, but also in the temperature range where PNIPAM is soluble. It is noteworthy that this effect of ethanol molecules selection in the first solvation shell occurs in a binary mixture where the cosolvent content is so low.

To characterize the interaction between polymer and water/ethanol molecules we have analysed how the number of PNIPAM-water/ethanol hydrogen bonds changes with temperature (Figures~\ref{fgr:solvent}(b) and~\ref{fgr:solvent}(c)). It is important to note that each PNIPAM repeating unit can form 3 HBs in the fully hydrogen bonded state: 2 HBs with the carbonyl acceptor group and 1 HB with the NH donor group. The temperature dependence of the number of PNIPAM-water HBs, reported in Figure~\ref{fgr:solvent}(b), shows a sigmoidal profile similar to what observed in the simulations in aqueous solution~\cite{tavagnacco2018molecular} and predicted by Okada and Tanaka for water molecules bound to PNIPAM in the cooperative transition from soluble to insoluble states~\cite{okada2005cooperative}. However, the number of PNIPAM-water HBs in aqueous solution decreases from about 2.6 to 2.1 across the coil-to-globule transition temperature~\cite{tavagnacco2018molecular}, while the corresponding extreme values in the presence of ethanol molecules vary from 2 and 1.5. These findings indicate that, while a significant number of PNIPAM-water HBs is kept formed even in the binary mixture of solvents, a competition for PNIPAM hydrophilic sites between water and ethanol molecules occurs. In the case of PNIPAM-ethanol hydrophilic interactions, the number of PNIPAM-ethanol HBs is considerably lower than that of PNIPAM-water HBs (Figure~\ref{fgr:solvent}(c)) and it is not affected by the conformational transition. The HBs formed by PNIPAM with ethanol are not sufficient to compensate the loss of polymer-water HBs (Figures~\ref{fgr:solvent}(b) and~\ref{fgr:solvent}(c)). A greater number hydrogen bonds of this polymer with ethanol was found in simulations of PNIPAM-water-ethanol solution, but at a larger content of cosolvent~\cite{Backes2017}.

By considering the total number of HBs formed between the polymer and solvent molecules, our results show a decrease of PNIPAM hydrogen bonding capability in the binary mixture, as compared to water. This finding can be attributed to a hindered accessibility of water to amide groups, in presence of surrounding ethanol molecules, as postulated by Pica and Graziano~\cite{pica2016}. To fully characterize the composition of the polymer first solvation shell, we have then monitored the temperature dependence of the total number of water/ethanol molecules in this domain and we have also identified those surrounding PNIPAM hydrophilic or hydrophobic groups, as displayed in Figures~\ref{fgr:solvent}(d) and~\ref{fgr:solvent}(e). In the case of hydration water molecules, a net decrease of their average number takes place at the coil-to-globule transition temperature in proximity to both hydrophilic and hydrophobic PNIPAM moieties. Moreover, it can be noticed that the partition of water molecules between hydrophilic and hydrophobic polymer sites is independent on temperature, and hence on chain conformation, in the whole investigated temperature range, with about the 20\% of first shell water molecules surrounding hydrophilic groups. This reveals that the ratio between PNIPAM hydrophilic and hydrophobic sites accessible to water molecules is not affected by temperature. Differently from water, the average number of first shell ethanol molecules is constant with temperature in proximity of both hydrophilic and hydrophobic PNIPAM moieties (Figure~\ref{fgr:solvent}(e)), suggesting that the cosolvent affinity for PNIPAM does not change with temperature. The ethanol-PNIPAM interaction is characterized by about a 10\% of contacts with hydrophilic amide groups, the overwhelming amount of ethanol molecules interacting through contacts with hydrophobic isopropyl groups. Finally, the results of Figures~\ref{fgr:solvent}(d) and~\ref{fgr:solvent}(e) demonstrate that the increase of ethanol concentration in the first solvation shell, detected at temperatures above the transition (Figure~\ref{fgr:solvent}(a)), is attributable to the depletion of water molecules from the surrounding of both hydrophilic and hydrophobic regions of the polymer.

\subsection{PNIPAM segmental mobility and solvents dynamics}

\begin{figure}[H]
\centering
\includegraphics[width=12 cm]{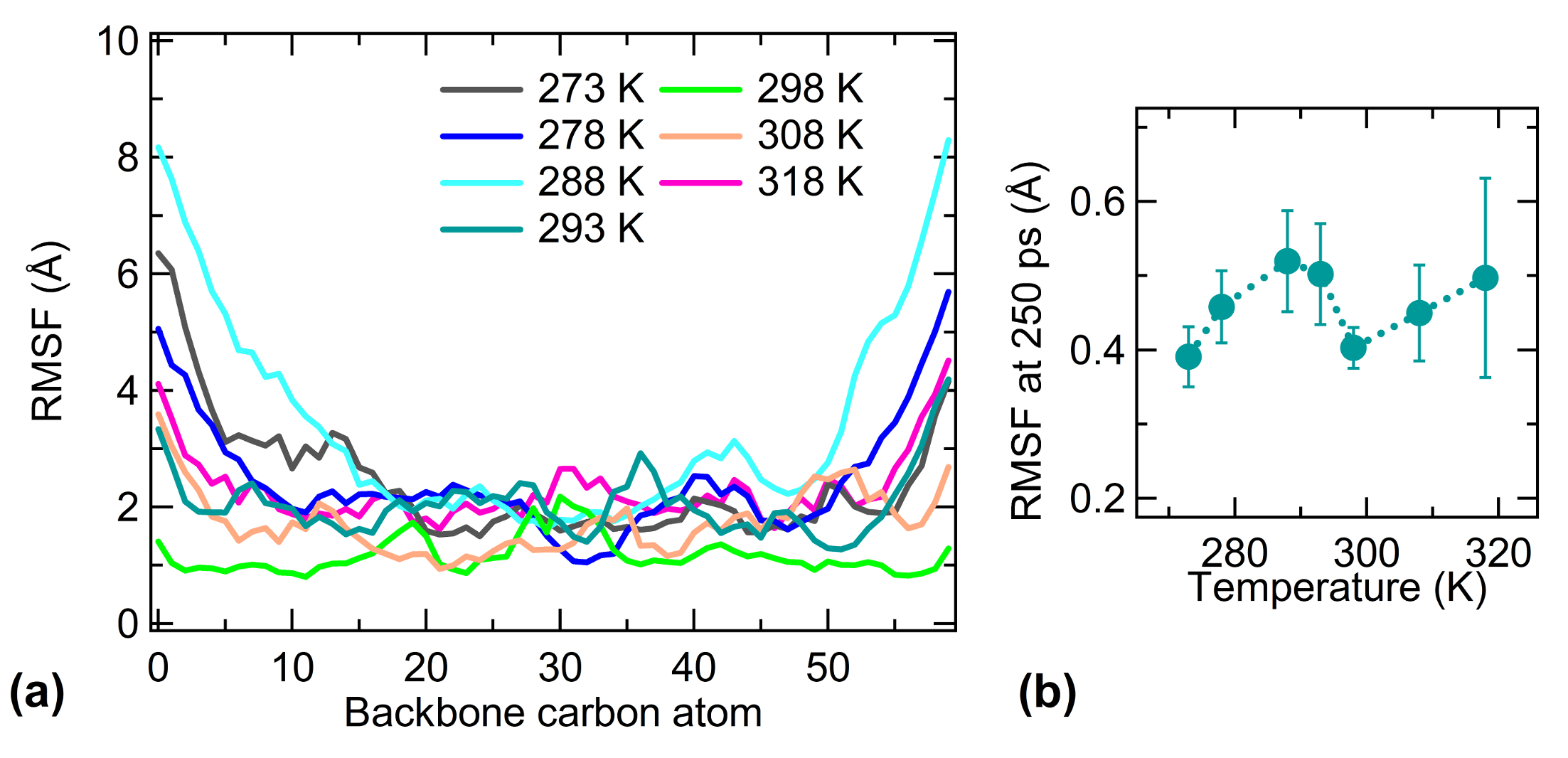}
\caption{(\textbf{a}) Root mean square fluctuations of PNIPAM backbone carbon atoms calculated at the different temperatures investigated. The abscissa value is the number of backbone carbon atom along the chain. (\textbf{b}) RMSF of PNIPAM backbone carbon atoms averaged over 250 ps, not including the first and last five residues of the polymer chain. The dotted line is a guide to the eye.}
  \label{fgr:dynpnipam}
\end{figure}

The coil-to-globule transition involves structural rearrangements of the PNIPAM chain which can influence the polymer internal flexibility. We have investigated the effect of the conformational transition on the polymer segmental mobility by calculating the root mean squared fluctuations of the backbone carbon atoms as a function of temperature, as reported in Figure~\ref{fgr:dynpnipam}(a). Typically, residues located in the chain ends exhibit a higher mobility than the internal ones, which is consistent with their lower topological constraints. In particular, the enhancement of the dynamics of the external residues, detected at 288 K, is related to the population of extended conformations with significant structural rearrangements, as displayed by the time evolution of the radius of gyration at this temperature (Figure~\ref{fgr:pnipam}(a)). The coil-to gobule transition entails a reduction of internal mobility and this is particular evident at 298 K, where the RMSF of backbone carbon atoms reaches the lowest values.

A further description of the local dynamics of PNIPAM atoms can be provided by averaging the root mean square fluctuations of backbone carbon atoms over a time interval comparable to that accessible by using experimental approaches, e.g. quasi-elastic neutron scattering experiments. Figure~\ref{fgr:dynpnipam}(b) shows the RMSF of backbone carbon atoms averaged over 250 ps and not including the first and last five chain residues. The polymer mobility increases monotonically with temperature, with a sudden reduction at the transition temperature, as experimentally observed for PNIPAM in the water-methanol solution with cosolvent molar fraction of $x_{ET}\cong0.15$ ~\cite{kyriakos2016}. Overall, the behavior of PNIPAM segmental mobility in the binary mixture of solvents is qualitatively similar to the one observed in aqueous solution~\cite{tavagnacco2018molecular}. This suggests that the polymer conformational fluctuations do not play a major role in the co-nonsolvency effect of ethanol, as instead it was found in the case of PNIPAM-water-methanol systems~\cite{rodriguez2015mechanism}.

\begin{figure}[H]
\centering
\includegraphics[width=12 cm]{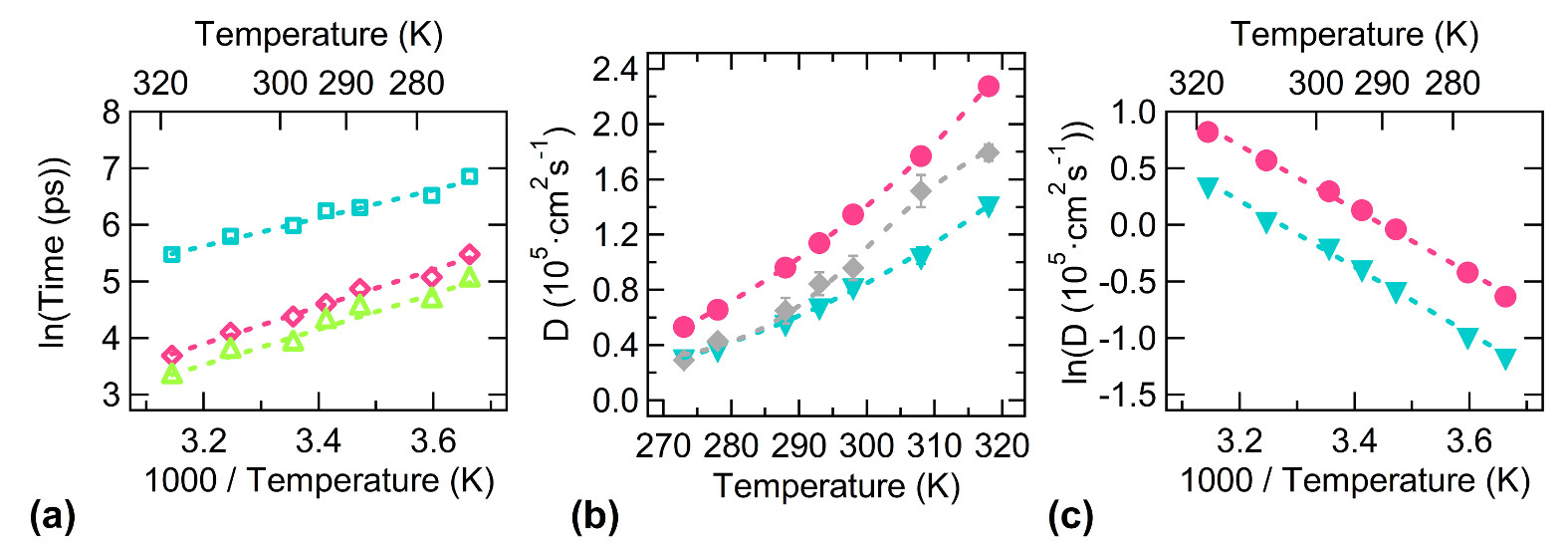}
\caption{(\textbf{a}) Comparison between the Arrhenius plot of the residence time in the first solvation shell of water molecules (magenta empty diamonds) and ethanol molecules (cyan empty squares) and the lifetime of PNIPAM-water hydrogen bonds (green empty triangles). (\textbf{b}) Temperature dependence of self diffusion coefficient as calculated for bulk water (magenta full circles), hydration water (gray full diamonds) and bulk ethanol (cyan full triangles). Dashed lines are guide to the eye. (\textbf{c}) Arrhenius plot of the self diffusion coefficient of water (magenta full circles) and ethanol (cyan full triangles) molecules in the bulk mixture.}
  \label{fgr:dynsolvent}
\end{figure}

In order to investigate the influence of the polymer and the effect of the cosolvent on water and ethanol dynamics, we have first monitored the residence time of ethanol and water molecules in the first solvation shell of PNIPAM. The autocorrelation functions reported in Figure S2 and the corresponding residence times summarized in Table S2 of the Supplementary Material reveal that the time of exchange of first solvation shell ethanol molecules with the more external solvent molecules is higher as compared to that calculated for water molecules, due to the slower diffusion motion of larger molecules. Overall, the values of residence time of water molecules in the binary mixture of solvents are comparable to those obtained from simulation studies in aqueous solution below and above the coil-to-globule transition temperature~\cite{chiessi2016}. In addition, water and ethanol residence times follow an Arrhenius behavior (see Figure~\ref{fgr:dynsolvent}(a)) with an activation energy of $27(\pm1)$~kJ~mol$^{-1}$ and $21(\pm1)$~kJ~mol$^{-1}$, respectively. The former value is in quantitative agreement with the activation energy for the exchange of first shell water molecules estimated from dielectric relaxation experiments on PNIPAM chains in aqueous solution, which is equal to $25$~kJ~mol$^{-1}$~\cite{ono2006hydration}. In Figure~\ref{fgr:dynsolvent}(a) it is also displayed the Arrhenius plot of the lifetime of PNIPAM-water HBs which is characterized by an activation energy of $26(\pm2)$~kJ~mol$^{-1}$, equal to that obtained for the exchange of first shell water molecules. Therefore, water mobility in the surrounding of the polymer is mainly dictated by this HB interaction. In the case of ethanol, the lifetime of PNIPAM-ethanol hydrogen bonding interactions cannot be estimated, due to the scarce number of HBs. However, the lower value of activation energy for the exchange of first shell ethanol molecules, as compared to that of water molecules, suggests that the nature of the interaction with PNIPAM is different, as it is also shown by the higher relative number of contacts with hydrophobic groups for ethanol in the first solvation shell of the polymer (Figures~\ref{fgr:solvent}(d) and~\ref{fgr:solvent}(e)).

We have further explored the solvent dynamical properties by calculating the self diffusion coefficient of water and ethanol molecules. Figure~\ref{fgr:dynsolvent}(b) compares the temperature dependence of the diffusion coefficient of water and ethanol molecules in the bulk mixture and of hydration water molecules. At 298 K, the bulk diffusion coefficients of water and ethanol molecules obtained from our simulations are $D_w=1.35(\pm0.03)\cdot10^{-5}~cm^2~s^{-1}$ and $D_E=0.83(\pm0.03)\cdot10^{-5}~cm^2~s^{-1}$, in good agreement with the experimental values of $D_w=1.249(\pm0.003)\cdot10^{-5}~cm^2~s^{-1}$ and $D_E=0.732(\pm0.003)\cdot10^{-5}~cm^2~s^{-1}$ measured at an ethanol molar fraction of $x_{ET}=0.1$~\cite{par2013mutual}. Figure~\ref{fgr:dynsolvent}(b) shows that the presence of PNIPAM has a strong influence on water mobility determining a significant reduction of the water self diffusion coefficient with a sigmoidal dependence on temperature, as previously detected in experimental and computational studies on PNIPAM aqueous solutions~\cite{osaka2009quasi,tavagnacco2018molecular}. As discussed for the lifetime of hydrogen bonding interactions, the diffusion coefficient of hydration ethanol molecules cannot be estimated. However, the temperature dependence of the mean squared displacement calculated at 250 ps shows that ethanol mobility is also reduced in the first solvation shell (Figure S3 of the Supplementary Material).

Both bulk $D_w$ and $D_E$ follow an Arrhenius behavior (see Figure\ref{fgr:dynsolvent}(c)) with similar activation energies of $23(\pm1)$~kJ~mol$^{-1}$ and $24(\pm1)$~kJ~mol$^{-1}$, respectively. Such a result suggests that the perturbation of the same kind of interactions is involved in the diffusion of the two solvents. We can attribute this role to the hydrogen bonding between water molecules. Indeed, being water the majority component, the diffusion motion of ethanol implies the concomitant diffusion of surrounding water molecules, which is governed by their hydrogen bonding. This hypothesis is confirmed by the value of water-water HB activation energy for molecules in the bulk mixture, equal to $25(\pm1)$~kJ~mol$^{-1}$ (Figure S4 of the Supplementary Material). It is noteworthy that the activation energy for the diffusion of pure liquid ethanol is lower than that for pure water, experimentally evaluated as equal $17.4$~kJ~mol$^{-1}$~\cite{hurle1985self,meckl1988self} and $21.5$~kJ~mol$^{-1}$~\cite{holz2000temperature,easteal1989diaphragm}, respectively. However, such activation energies were found to become similar in the simulation of the binary mixture~\cite{pothoczki2018temperature}.

\subsection{Molecular mechanism of the coil-to-globule transition in water/ethanol mixture}

In the context of our model, the phenomenon of co-nonsolvency consists of the formation of a stable globule state for PNIPAM at 293 K, a temperature where the polymer is soluble (namely it mainly assumes coil conformations) if dissolved in water or in ethanol. The coil-to-globule transition occurs with a discontinuity in the extent of the polymer interaction with water, as shown in Figures~\ref{fgr:solvent}(b) and~\ref{fgr:solvent}(d). The same discontinuity is also observed in aqueous solution, with a similar reduction of the number of polymer-water HBs and of water molecules in the surrounding of hydrophobic groups~\cite{tavagnacco2018molecular}. However, why is the transition temperature anticipated of about 10 K in the mixture containing, on average, 1 ethanol molecule out of 13 water molecules? A first finding from our simulations is that the mixture environment surrounding PNIPAM in proximity of the transition temperature is actually composed by 1 ethanol molecule out of 6.6 water molecules despite a much reduced overall concentration in the mixture, which reinforces the role of the cosolvent. Taking into account the volume difference between water and ethanol molecules, one third of the polymer-solvents interface is thus approximately occupied by the alcohol, with the consequence of a decrease of the hydration degree of PNIPAM in the coil state. This picture is in agreement with the interpretation of co-nonsolvency by Pica and Graziano~\cite{pica2016}, which recognises the rational of this phenomenon in the competition between water and cosolvent molecules to establish attractive interactions with the PNIPAM surface.

\begin{figure}[H]
\centering
\includegraphics[width=12 cm]{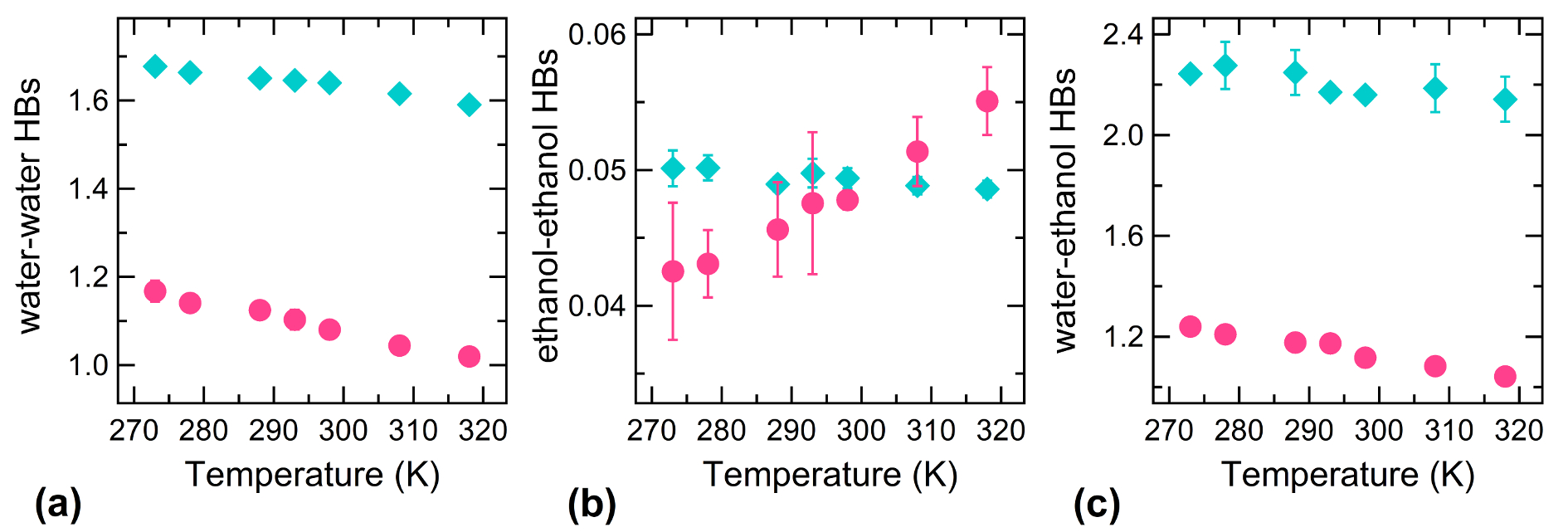}
\caption{Temperature dependence of the average number of hydrogen bonds between water molecules (\textbf{a}), ethanol molecules (\textbf{b}) and water-ethanol molecules (\textbf{c}) normalized to the number of water molecules (\textbf{a}) and ethanol molecules (\textbf{b}) and (\textbf{c}). Data are reported for first shell solvent molecules (magenta circles) and bulk solvent molecules (cyan diamonds). When not distinguishable, error bars are within symbol size.}
  \label{fgr:HBs}
\end{figure}

The relevant conformational rearrangement of the polymer occurring at the transition could be related to modifications of the structure of the first solvation shell, stabilizing the globule conformation at lower temperatures, as compared to the aqueous solution. To clarify the influence of the structuring of the solvation shell, we have evaluated how water-cosolvent interactions change across the coil-to globule transition both in proximity of PNIPAM and in the bulk of the mixture. Figures~\ref{fgr:HBs}(a-c) report the calculated number of HBs between solvent molecules of the same type and between the two different solvent molecules. The temperature dependence of the water-cosolvent interactions reveal that: i) the number of water-water HBs is lower in the first solvation shell than in the bulk of solution due to the interface with the polymer and it decreases linearly with temperature; ii) the number of ethanol-ethanol HBs is almost constant with temperature in the bulk of solution, whereas in the first solvation shell for temperatures below 293 K the number of HBs is reduced compared to that in the bulk and it becomes higher than the bulk composition for temperatures above 293 K; and iii) the number of water-ethanol HBs is lower in the first solvation shell than in the bulk of solution, as observed for water-water HBs, and it is significantly higher than the number of PNIPAM-ethanol HBs (Figure~\ref{fgr:solvent}(b)). These findings suggest that ethanol molecules prevalently interact with the polymer chain through dispersion interactions and that the chain collapse favours the formation of ethanol-ethanol contacts. It is noteworthy that the decreasing behaviour of the number of HBs as a function of temperature is a consequence of the exothermic character of this interaction. The exception observed in Figure~\ref{fgr:HBs}(b), concerning inter-ethanol HBs in the first solvation shell, derives from the increase of local ethanol concentration, concomitant to the coil-to-globule transition. For the same reason, namely the depletion of water molecules, the slope of the temperature behaviour of water-water HBs in the first solvation shell has a larger absolute value, as compared to the corresponding value for water in the bulk mixture (Figure~\ref{fgr:HBs}(a)). However, no discontinuity is detected across the transition for these HBs interactions, which suggests to exclude them between the factors determining the decrease of the coil-to-globule transition temperature in the presence of ethanol.

\begin{figure}[H]
\centering
\includegraphics[width=10 cm]{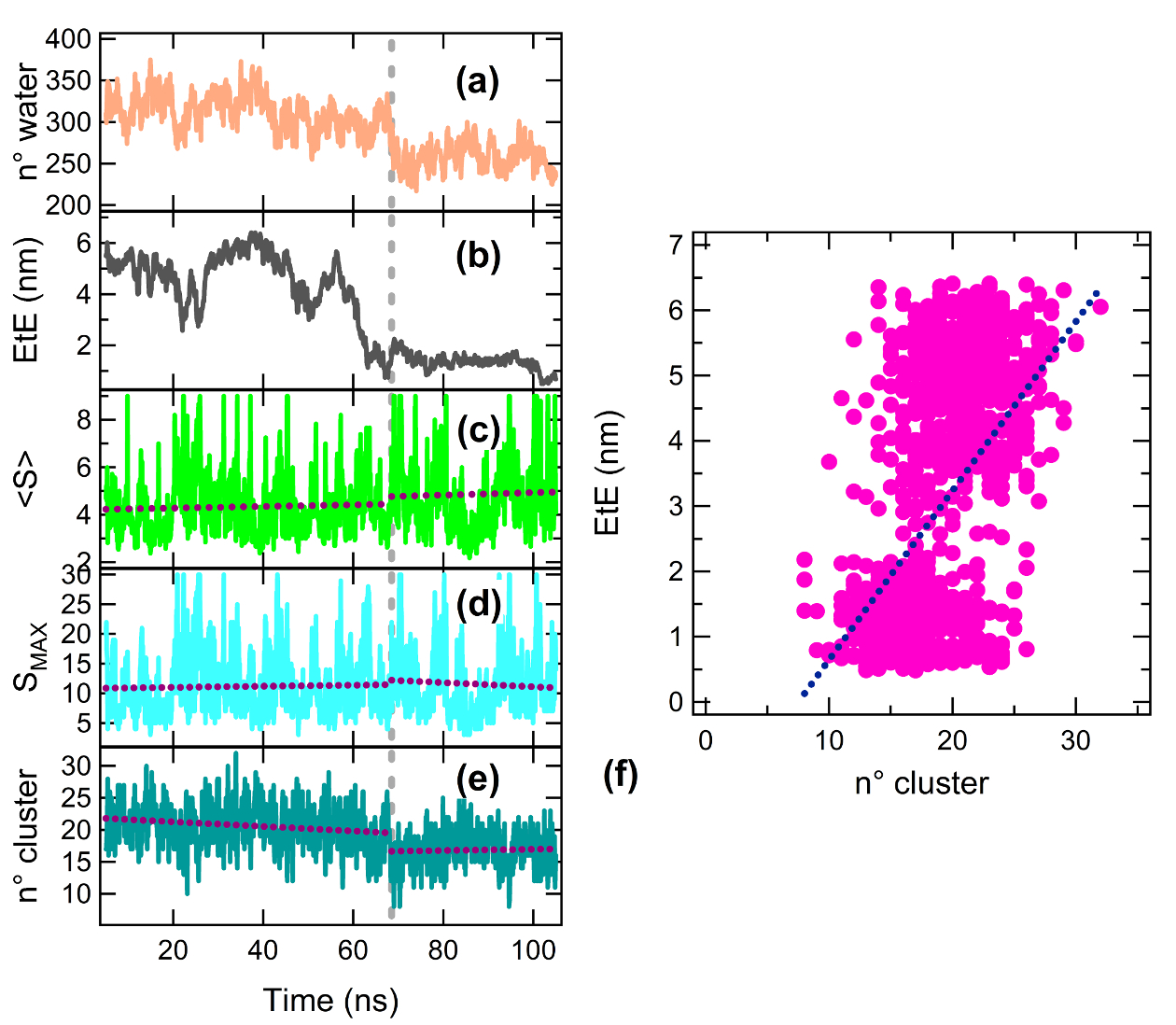}
\caption{Time evolution from the initial 105 ns of the simulation at 293 K of (\textbf{a}) the number of first shell water molecules (orange), (\textbf{b}) the end to end distance ($EtE$) of the PNIPAM chain (gray), (\textbf{c}) the average size ($<S>$) of the ethanol clusters in the first solvation shell (green), (\textbf{d}) the maximum size ($S_{MAX}$) of the ethanol clusters in the first solvation shell (light blue) and (\textbf{e}) the number of ethanol clusters in the first solvation shell (dark blue). Vertical dashed lines indicate the time at which the conformational transition has occurred and horizontal dotted lines are linear fit of the data. (\textbf{f}) Correlation between the end-to-distance of the PNIPAM chain and the number of ethanol clusters in the first solvation shell from the initial 105 ns of the simulation at 293 K. The dotted line is the linear fit of the data.}
  \label{fgr:clu}
\end{figure}

It was postulated that cosolvent molecules, adsorbed on distant regions of the polymer chain, can favour the formation of intra-chain contacts and hence contribute to facilitate the coil-to-globule transition~\cite{mukherji2014}. We have therefore monitored the formation and evolution of clusters of ethanol molecules in the first solvation shell during the conformational transition, to clarify the role of the cosolvent in the mechanism of the process. In particular, we have considered the simulation data collected at 293 K, a temperature at which the chain collapse clearly takes place in the first 70 ns. By calculating the time evolution of the number of first shell water molecules and of the end to end distance of the polymer chain (Figure~\ref{fgr:clu}(a and b)) we have identified that a net variation of these observables occurs at 68.5 ns. Figures ~\ref{fgr:clu}(c, d and e) display the average size, the maximum size and the total number of clusters of ethanol molecules in the first solvation shell of the polymer chain during the same time interval. At times longer than 68.5 ns, the average and the maximum cluster size of ethanol molecules increase, while the total number of cluster is reduced. These behaviours indicate that the conformational transition occurs with a significant loss of water molecules (Figure~\ref{fgr:clu}(a)) that promotes the formation of a lower number of larger ethanol clusters. The correlation between the end-to-end distance of the polymer chain and the number of ethanol clusters is further highlighted in Figure~\ref{fgr:clu}(f). However, the analysis of the distribution of ethanol molecules at the interface with PNIPAM shows a uniform adsorption, hence a trigger of the coil-to-globule transition by the attraction between ethanol-rich segments of the chain seems unlikely. Nonetheless, this kind of mechanism could intervene for PNIPAM chains with higher degree of polymerization.

The free energy variation for the coil-to-globule transition, $\Delta G_{c-g}(T)$, is modulated by temperature according to:

\begin{eqnarray}
\Delta G_{c-g}(T)&=&\Delta H_{c-g}-T \cdot \Delta S_{c-g}
\label{Eq:Dw}
\end{eqnarray}

assuming as negligible the temperature dependence of the transition enthalpy and entropy, $\Delta H_{c-g}$ and $\Delta S_{c-g}$, respectively. In aqueous environment $\Delta H_{c-g}$ is positive, mainly for the reduction of PNIPAM-water interactions, consisting of both hydrogen bonds with amide groups and interactions with hydrophobic isopropyl groups~\cite{tavagnacco2018molecular}. The value of $\Delta S_{c-g}$ is overall positive for the decrease of excluded volume to solvent molecules, which leads to a concomitant increase of their translational entropy, notwithstanding the reduction of conformational entropy of the polymer. Therefore the transition temperature value, given by the $\Delta H_{c-g}$ / $\Delta S_{c-g}$  ratio, is affected by factors of both enthalpic and entropic nature. Experiments of differential scanning calorimetry detected a decrease of $\Delta H_{c-g}$ from 5.5 to about 3$~kJ~mol^{-1}$ of repeating units, moving from water to water-ethanol at $x_{ET} = 0.07$~\cite{Trappe2014}. Also the transition entropy, $\Delta S_{c-g}$, goes through a reduction in the water-ethanol mixture, as compared to the aqueous solution, because of the greater size of adsorbed ethanol molecules. The decrease of $\Delta H_{c-g}$, not compensated by a corresponding decrease of $\Delta S_{c-g}$ causes a reduction of the transition temperature value. The drop of PNIPAM-water HBs at the transition and the release of water from hydrophobic domains significantly contribute to the transition enthalpy, however the extent of polymer dehydration, of both hydrophilic and hydrophobic groups, at the transition is similar in water and water-ethanol solutions, as shown by our simulation results and by those in~\cite{tavagnacco2018molecular}. Moreover, the increase of ethanol clustering (Figure~\ref{fgr:clu}(c)) involves a very small net effect on enthalpy, for the association of apolar ethanol moieties. Lastly, enthalpic and entropic contributions related to structural rearrangements of the solvation shell at the transition are negligible (Figure~\ref{fgr:HBs}). On the basis of these considerations, our findings suggest that the lower enthalpy of water in the bulk of mixture, originated by the negative mixing enthalpy of the two solvents, plays a significant role on the reduction of $\Delta H_{c-g}$ in water-ethanol solution and hence on the decrease of the transition temperature~\cite{Trappe2014}.

\section{Conclusions}

In the present study we explore the molecular mechanism of the coil-to-globule transition of a PNIPAM linear chain in a mixture of water/ethanol at a low alcohol molar fraction by using all-atom molecular dynamics simulations. Our framework successfully reproduces the conformational rearrangement at a temperature of $\sim$289 K, in full agreement with the experimental findings~\cite{Trappe2014}. We observe that the presence of the cosolvent does not affect the formation of PNIPAM intra-molecular hydrogen bonds, but it decreases the number of PNIPAM-water hydrogen bonds with respect to the pure aqueous solution system~\cite{tavagnacco2018molecular}. However, this variation is not compensated by the formation of PNIPAM-ethanol hydrogen bonds, whose number appears also scarcely dependent on temperature. These features reveal the presence of a competition between water and ethanol molecules to interact with the polymer, as it would be expected if geometric frustration played a driving role in the chain collapse~\cite{pica2016}. Furthermore, we find that the coil-to globule transition occurs with a major loss of water molecules located in proximity to isopropyl groups and that the molar fraction of ethanol molecules in the first solvation shell of PNIPAM is higher than in the bulk of solution and it grows with temperature. Our results also show that PNIPAM segmental mobility in the binary mixture of solvents is qualitatively similar to the one observed in aqueous solution, thus revealing that the polymer conformational fluctuations cannot be identified as the main responsible of the co-nonsolvency effect in the presence of ethanol. In addition, we reproduce the experimental self diffusion coefficients of the binary mixture of solvents in the bulk of solution and we find a significant slowing down of the self diffusion coefficient of hydration water molecules and of ethanol mobility in the surrounding of polymer. A detailed analysis of the water-cosolvent interactions in the first solvation shell and in the bulk of solution indicates that ethanol molecules predominantly interact with PNIPAM through van der Waals interactions and that the coil-to-globule transition involves an increase of aggregation of ethanol molecules in the first solvation shell of the polymer.

Our findings show a localization of cosolvent molecules at the polymer interface, but the calculated activation energies for the processes of exchange of first solvation shell water/ethanol molecules reveal that the interaction of ethanol with PNIPAM is more labile than that of water. Therefore our results do not recognize the preferential affinity of ethanol to PNIPAM as the primary responsible for the decrease of the LCST, as reported for water-methanol mixtures~\cite{mukherji2013}. Furthermore, we observe that the mechanism of the coil-to globule transition in water-ethanol mixture involves a change in the composition of water in the first solvation shell similar to the one detected in aqueous solution~\cite{tavagnacco2018molecular}, and no discontinuity is detected in the structuring of the first solvation shell. These characteristics support the idea that the major reduction of molar partial enthalpy of water for the mixing with ethanol, and the consequent decrease of the water chemical potential in the bulk of the mixture, are key factors for the anticipation of the transition temperature~\cite{Trappe2014}.

Finally, we believe that this contribution can stimulate further experimental characterization of PNIPAM co-nonsolvency in water-ethanol mixtures, which, as compared to the PNIPAM-water-methanol ternary system, is still quite unexplored.

\section*{Appendix A. Supplementary data}{Supplementary data to this article can be found online; Table S1: Density of the model solutions; Figure S1: Definition of first solvation shell; Figure S2-S3-S4 and Table S2: Dynamical behavior of water and ethanol molecules.}

\section*{Acknowledgments}
The research leading to these results has received funding from the European Research Council - ERC MIMIC (ERC-CoG-2015, Grant No. 681597) and from MIUR FARE SOFTART (R16XLE2X3L). Support for computational work by CINECA ISCRA grants is gratefully acknowledged.

\section*{Copyright}
$\copyright$2019. This manuscript version is made available under the CC-BY-NC-ND 4.0 license http://creativecommons.org/license/by-nc-nd/4.0/


\newpage
\renewcommand{\thefigure}{S\arabic{figure}}
\renewcommand{\thetable}{S\arabic{table}}
\setcounter{figure}{0}

\section*{Supplementary Material}

\section*{S1. Density of the model solutions}

The density of the model solution as a function of the temperature was calculated over the last 40 ns of trajectory. The comparison between the density values obtained from simulations and the experimental ones of the mixture of ethanol and water at $x_{ET} = 0.07$ is reported in Table~\ref{tbl:density}. The discrepancy between simulation and experimental values is below 1\%.
\\

\begin{table}[H]
  \centering
  \caption{\textbf{Density of the system} }
  \label{tbl:density}
  \begin{threeparttable}
  \setlength{\tabcolsep}{24pt}
  \begin{tabular}{ccc}
   \hline
    T	    & $\rho$                    & $\rho^*$   \\
    (K) 	& (g$\cdot cm^{-3}$)        & (g$\cdot cm^{-3}$) \\
   \hline
   273		& 0.9740 ($\pm$0.0001)      & - \\
   278		& 0.9724 ($\pm$0.0001)      & - \\
   288		& 0.9682 ($\pm$0.0001)      & 0.97433 \\
   293		& 0.9659 ($\pm$0.0001)      & 0.97259 \\
   298		& 0.9633 ($\pm$0.0001)      & 0.97062 \\
   308		& 0.9577 ($\pm$0.0001)      & 0.96607 \\
   318		& 0.9514 ($\pm$0.0001)      & - \\
   \hline
   \end{tabular}
        \begin{tablenotes}
        \footnotesize
        \item \emph{$T$ is the temperature, $\rho$ is the density of the model solution and $\rho^*$ is the experimental density of water-ethanol mixtures from reference~\cite{density1913}.}
        \end{tablenotes}
    \end{threeparttable}
\end{table}

\section*{S2. Definition of first solvation shell}

\begin{figure}[H]
\centering
\includegraphics[width=11 cm]{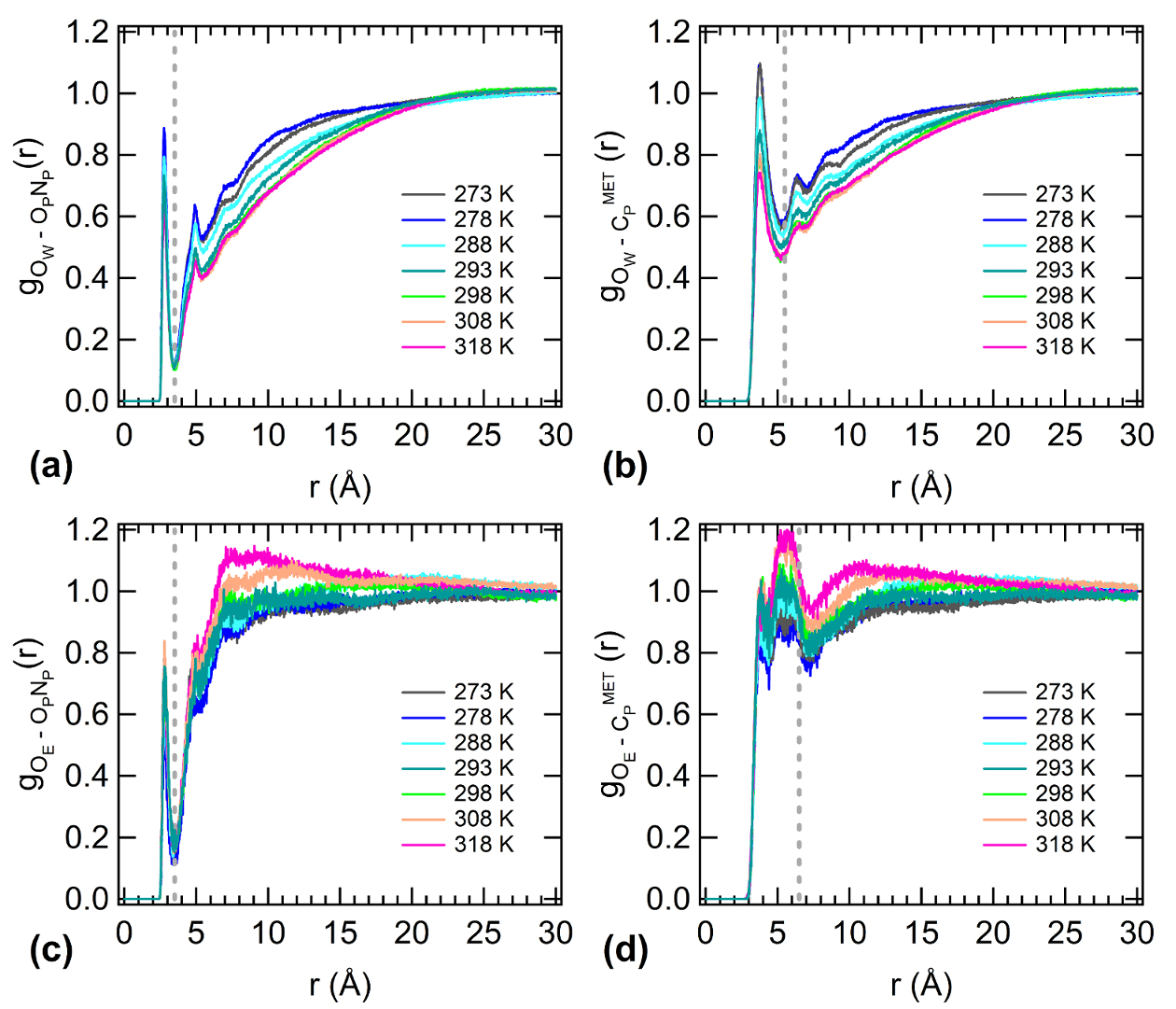}
\caption{Radial distribution function of (\textbf{a})/(\textbf{c}) water/ethanol oxygen atoms $O_W$/$O_E$ around PNIPAM nitrogen $N_P$ or oxygen atoms $O_P$ and (\textbf{b})/(\textbf{d}) water/ethanol oxygen atoms $O_W$/$O_E$ around PNIPAM methyl carbon atoms $C_P$ used to define the cutoff values of the first solvation shell. Data calculated at 273 K, 278 K, 288 K, 293 K, 298 K, 308 K, and 318 K are shown in dark gray, blue, cyan, dark green, green, orange and magenta, respectively. The vertical dashed lines indicate the cutoff values employed in this work.}
  \label{fgr:gdr}
\end{figure}

The first solvation shell of the single chain was defined by calculating the radial distribution functions of solvent molecules around PNIPAM hydrophilic and hydrophobic groups (Figure~\ref{fgr:gdr}(a-d)). The positions of the first minima of the radial distribution functions were taken as cutoff distances of the first solvation shell.

\section*{S3. Dynamical behavior of water and ethanol molecules}
\begin{figure}[H]
\centering
\includegraphics[width=12 cm]{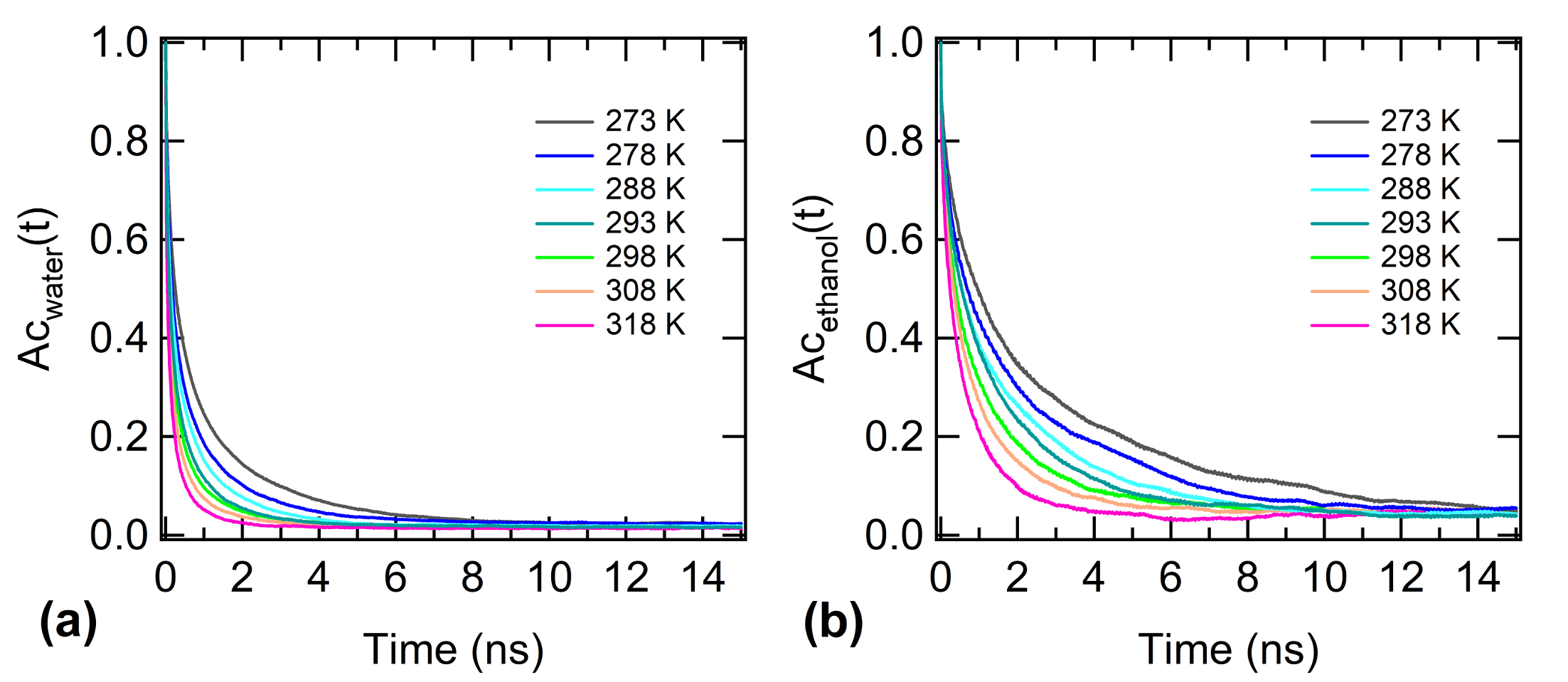}
\caption{Time behavior of the number fraction of water (\textbf{a}) and ethanol (\textbf{b}) molecules residing in the first solvation shell as a function of the temperature. Data obtained at 273 K, 278 K, 288 K, 293 K, 298 K, 308 K, and 318 K are shown in dark gray, blue, cyan, dark green, green, orange and magenta, respectively.}
  \label{fgr:AC}
\end{figure}

\begin{figure}[H]
\centering
\includegraphics[width=6.5 cm]{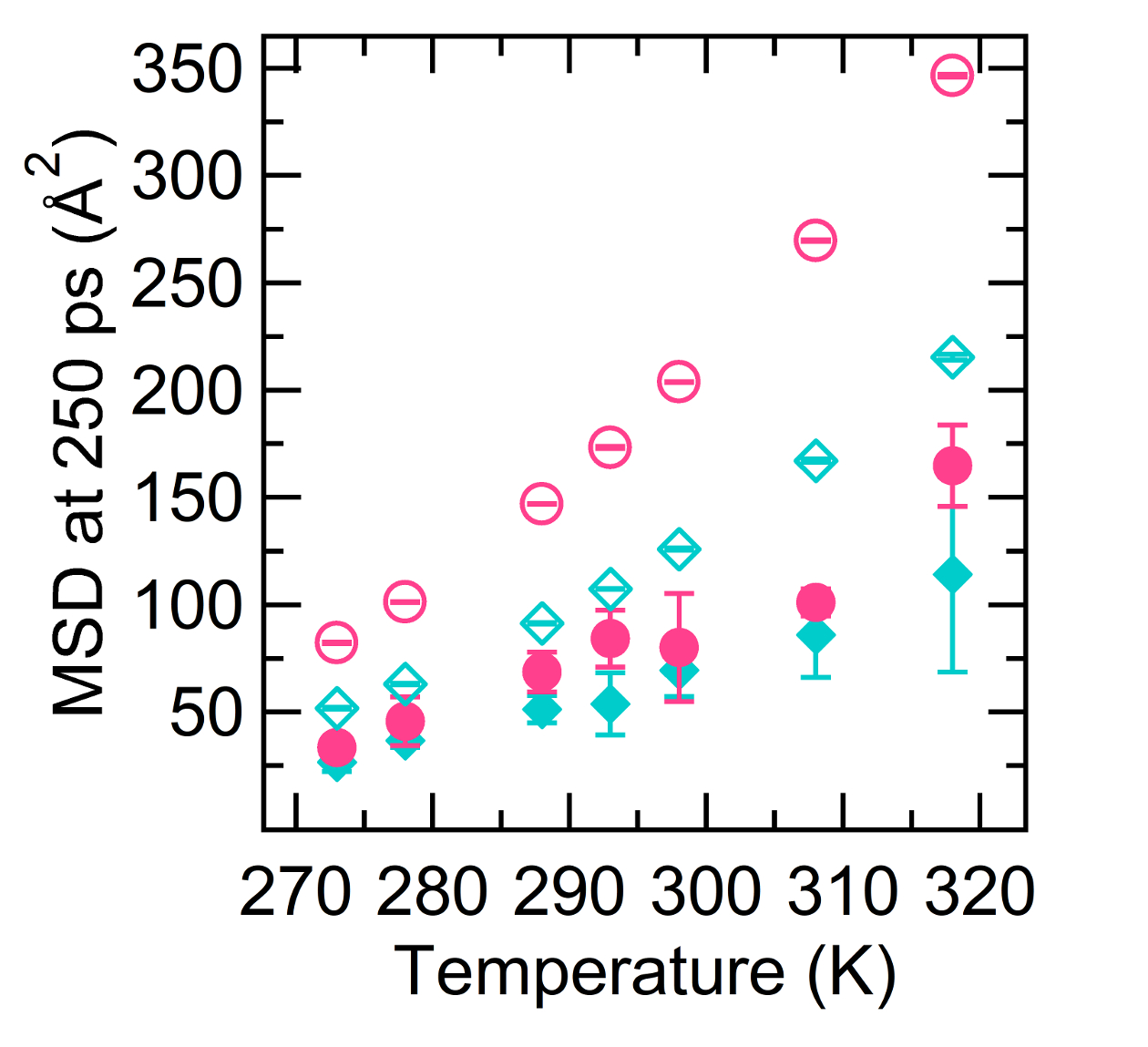}
\caption{Temperature dependence of the mean squared displacement calculated at 250 ps for hydration water/ethanol molecules (magenta full circles/cyan full diamonds) and bulk water/ethanol molecules (magenta empty circles/cyan empty diamonds).}
  \label{fgr:MSD}
\end{figure}

\begin{figure}[H]
\centering
\includegraphics[width=6.5 cm]{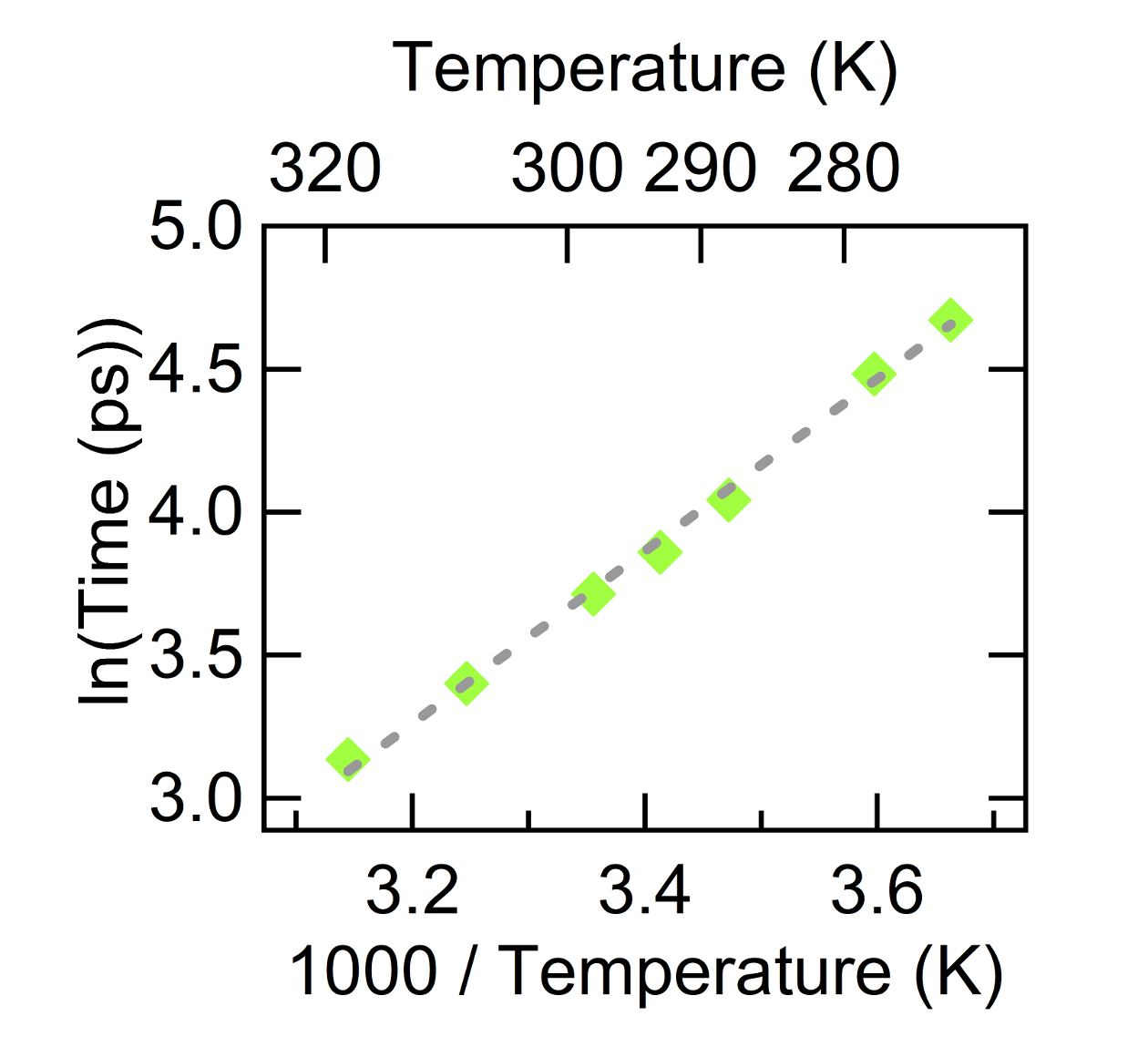}
\caption{Arrhenius plot of the lifetime of water-water hydrogen bonds in the bulk of the mixture.}
  \label{fgr:Bulk}
\end{figure}

The effect of PNIPAM on the solvent dynamics was studied by calculating the time evolution of the number fraction of water and ethanol molecules in the first solvation shell, as reported in Figure~\ref{fgr:AC}(a,b). The residence time of solvent molecules in the first solvation shell was estimated as the time at which the autocorrelation function is decayed of the 50\% of its amplitude (Table~\ref{tbl:time}). A further characterization of the solvent dynamics was carried out by comparing the temperature dependence of the mean squared displacement of hydration and bulk solvent molecules at 250 ps, as shown in Figure~\ref{fgr:MSD}. The lifetime of water-water hydrogen bonds in the bulk of the mixture was also evaluated as the time at which the autocorrelation function is decayed of the 90\% of its amplitude (Figure~\ref{fgr:Bulk}). These cutoff values were used in order to reduce the statistical noise. The activation energy for the hydrogen bonding between water molecules in the bulk of the mixure was calculated from Figure~\ref{fgr:Bulk}.

\begin{table}
  \centering
  \caption{\textbf{Solvent dynamics} }
  \label{tbl:time}
  \begin{threeparttable}
  \setlength{\tabcolsep}{12pt}
  \begin{tabular}{cccc}
   \hline
    T	& $\tau_{r,water}$ & $\tau_{r,ethanol}$ & $\tau_{PW}$ \\
    (K) 	& (ps)	& (ps)  & (ps) \\
   \hline
   273		& 247 ($\pm$1)			& 920 ($\pm$57) 		& 152 ($\pm$2)\\
   278		& 168 ($\pm$8)			& 690 ($\pm$85)			& 105 ($\pm$17)\\
   288		& 129 ($\pm$1)			& 540 ($\pm$42)			&  92 ($\pm$5)\\
   293		&  97 ($\pm$1) 			& 520 ($\pm$14)			&  72 ($\pm$4)\\
   298		&  79 ($\pm$3)			& 395 ($\pm$7)			&  50 ($\pm$4)\\
   308		&  59 ($\pm$1)			& 335 ($\pm$21)			&  43 ($\pm$4)\\
   318		&  43 ($\pm$1)			& 230 ($\pm$14)			&  27 ($\pm$4)\\
   \hline
   \end{tabular}
        \begin{tablenotes}
        \footnotesize
        \item \emph{$T$ is the temperature, $\tau_{r}$ is the residence time in the first solvation shell and $\tau_{PW}$ is the lifetime of PNIPAM-water hydrogen bonds.}
        \end{tablenotes}
    \end{threeparttable}
\end{table}

\end{document}